\documentclass[11pt,twoside]{article}
\usepackage{FUSE2004}
\usepackage{psfig}
\usepackage{epsf}
\usepackage{graphics}
\usepackage{lscape}
\def\edcomment#1{\iffalse\marginpar{\raggedright\sl#1\/}\else\relax\fi}
\reversemarginpar
\newcommand{\kms}{km\,s$^{-1}$}

\begin{document}
\title{Exploring Hot Gas in the Galactic Halo and High Velocity Clouds}
\author{Kenneth R. Sembach}
\affil{Space Telescope Science Institute, 3700 San Martin Dr., Baltimore, MD 21218}
\author{Bart P. Wakker, Blair D. Savage}
\affil{University of Wisconsin-Madison, 475 N. Charter St., Madison, WI 53706}
\author{Philipp Richter}
\affil{Institut fur Astrophysik und Extraterrestrische Forschung, 
Universit$\ddot{a}$t Bonn, Auf dem H$\ddot{u}$gel 71, 53121 Bonn, Germany}

\begin{abstract}
In the five years since its launch, the {\it Far Ultraviolet Spectroscopic
Explorer} has given the astronomical community a marvelous glimpse into
the realm of previously unexplored regions of the Milky Way and its
surroundings.  Absorption lines produced by gas along entire paths through 
the nearby universe have now been recorded in the spectra of distant
QSOs and active galactic nuclei.   Of the lines recorded, the 
\ion{O}{6} $\lambda\lambda 1031.926, 1037.617$ doublet is the best
tracer of highly ionized regions.  
Exciting discoveries include a definitive confirmation
of the hot Galactic halo postulated by Lyman Spitzer nearly 50 years
ago, the discovery of a hot, highly extended Galactic corona enveloping
the Milky Way out to distances of several tens of kiloparsecs, and
the discovery of an extensive network of highly ionized, high velocity 
clouds surrounding the Galaxy.
\end{abstract}
\thispagestyle{plain}

\section{Introduction}

It is a pleasure to be able to provide a review of FUSE 
\ion{O}{6} measurements in the Milky Way halo and high velocity cloud
system.  This work has been a major focus of the research efforts of 
many FUSE Team members over the past few years, and was the subject
of much planning and preparation before launch.  Before launch,
we had high expectations for the detection of \ion{O}{6} absorption
in the Galactic halo, and these came to fruition early in the 
mission.\footnote{The New York Times even reported the early Galactic halo
results in an article entitled ``Craft is Demystifying Milky Way's 
Sizzling Halo of Gas'' (J.N.\ Wilford, 13-Jan-2000).}
However, the 
large number of high-velocity clouds (HVCs) detected in \ion{O}{6} 
absorption came as a stunning surprise and was something that we had 
not expected prior to launch.  If anything, it is probably fair to say 
that many of us believed that with the exception of a few highly
ionized HVCs detected by HST/STIS, most of the high velocity gas was 
neutral, a view that has changed significantly since the FUSE 
observations.

\section{The FUSE \ion{O}{6} Survey}

We have conducted an extensive study of the \ion{O}{6} 
$\lambda\lambda1032,1038$ absorption
in the vicinity of the Milky Way as measured by FUSE for 100 AGN/QSO
sight lines and two distant halo star sight lines.  
This study was based on $\sim4$ Msec of exposure time, pulling together
data from many FUSE observing programs.  We show portions of the FUSE
spectra for two  sight lines in Figure~1.
The results of this survey have been published in a special issue of the 
Astrophysical Journal Supplement Series (May 2003, volume 146).  The
three papers in that issue include a description of the survey 
(Wakker et al.\ 2003), an investigation of the distribution and 
kinematics of \ion{O}{6} in the Galactic halo (Savage et al.\ 2003),
and an analysis of the highly ionized high velocity gas in the vicinity of 
the Galaxy (Sembach et al.\ 2003).  Rather than reference those 
papers repeatedly here, we simply draw upon the results therein
and refer the reader to those papers for further details and background
information.

\begin{figure}[h!]
\includegraphics{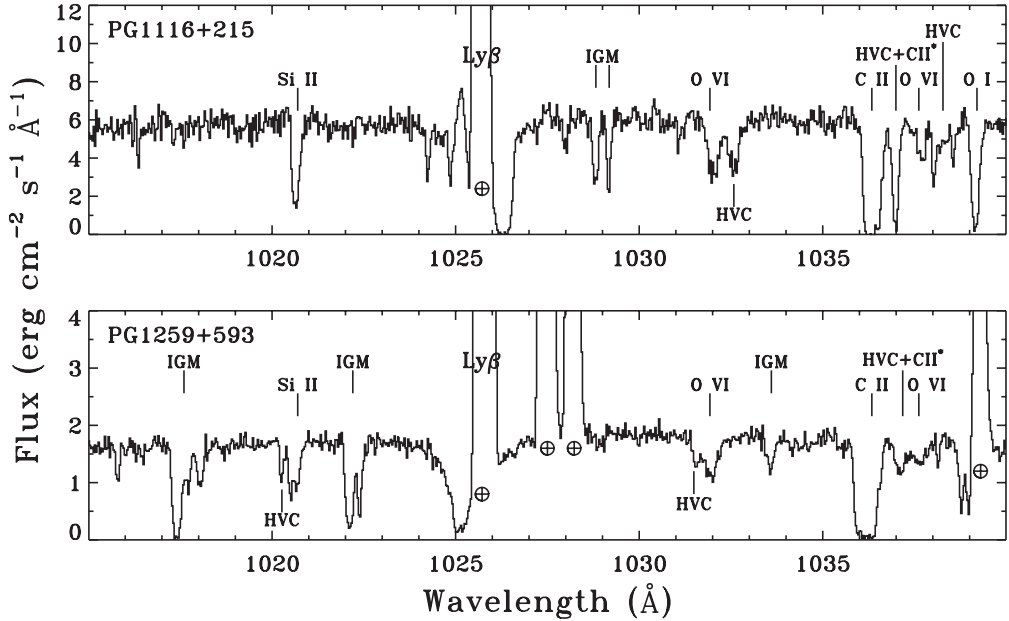}
\vspace{3.25in}
\caption{\footnotesize  Observations of two objects in the FUSE high velocity O\,{\sc vi} survey.  The data have 
a FWHM velocity resolution of $\sim20$ \kms.  
Low velocity Milky Way halo absorption features and high velocity 
absorption features are present along both sight lines.
Prominent interstellar lines, including the 
two lines of the O\,{\sc vi} $\lambda\lambda1031.926, 1037.617$ doublet,
are identified above each spectrum at their rest wavelengths.  
H\,{\sc i} and metal lines from 
intervening intergalactic clouds are marked in both panels.  
Unmarked absorption
features are interstellar H$_2$ lines.
Crossed circles mark the locations of terrestrial airglow emission
lines of H\,{\sc i}  and O\,{\sc i} . }
\end{figure}

Highlights of several
investigations of individual HVCs in the FUSE survey are summarized
in the papers by Fox et al.\ and Collins et al.\ in this volume.
Brief descriptions of FUSE observations of \ion{O}{6}
in the Milky Way disk and the Magellanic Clouds can be found in 
accompanying articles by Bowen et al.\ and Howk, respectively.

\subsection{Velocities}

With a few exceptions, we define \ion{O}{6} absorption in the survey 
as ``high velocity'' if it has $|v_{LSR}| \ga 100$ \kms, while we
attribute lower velocity gas to the thick disk and halo of the Milky Way.  
(For simplicity, we will use the term ``halo'' to denote both the thick 
disk and halo of the Milky Way.) The 
velocity distribution of the \ion{O}{6} features is shown in Figure~2.

\begin{figure}[ht!]
\includegraphics{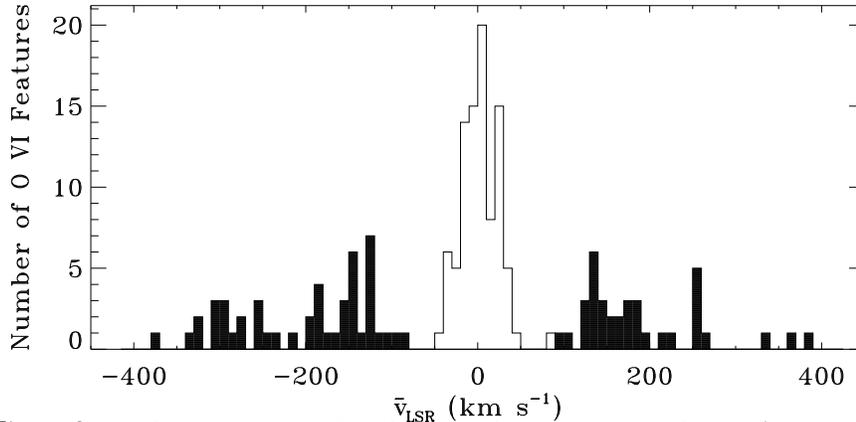}
\vspace{2.1in}
\caption{\footnotesize
Histograms of the O\,{\sc vi} line centroids for the Galactic halo 
(unshaded) and high
velocity gas (shaded). The bin size is 10 \kms.}
\end{figure}

The imprint of Galactic rotation can be seen in some of the absorption
profiles in low latitude ($|b| < 30\deg$) directions.   At higher
latitudes ($|b| \ga 45\deg$), outflows and inflows of gas in the 
halo are evident.  
The mean \ion{O}{6} velocities of the halo gas 
along high Galactic latitude sight lines range from --46 to +82 \kms\,
with a sample average of $0\pm21$ \kms.  This result holds for both
northern and southern Galactic latitudes, suggesting that the 
gas in both hemispheres is moving toward and away from the Galactic plane 
with approximately equal frequency.  

The high velocity \ion{O}{6} features have velocity centroids ranging 
from $-372 \le v_{LSR} \le -90$ km~s$^{-1}$ to 
$+93 \le v_{LSR}
\le +385$ km~s$^{-1}$. There are an additional six
confirmed or very likely ($>90$\% confidence) detections and two tentative 
detections of \ion{O}{6} 
between $v_{LSR} = +500$ and +1200 km~s$^{-1}$; these very high velocity
features probably trace intergalactic gas beyond the Local Group. 
By definition, the high velocity \ion{O}{6} features have velocities 
incompatible with those of Galactic rotation.  

\subsection{Column Densities}

For the 20 sight lines in the survey where both lines of the \ion{O}{6} 
doublet arising in the halo gas can be
measured reliably, the \ion{O}{6} column density inferred from direct
integration of the apparent column density profiles of the 
two lines is in good agreement in most cases.  A few exceptions exist
(see Table~2B in
Savage et al.\ 2003), but in general the excellent agreement indicates that
the more easily measured $\lambda1031.926$ line provides
a reliable estimate of the column density.  
Blending with other interstellar and 
intergalactic features is an important consideration for all sight lines; 
see Wakker et al.\ (2003) for details.

The Galactic halo \ion{O}{6} features in the survey have 
logarithmic 
column densities (cm$^{-2}$) of 13.85 to 14.78 with an average of 
$\langle \log N \rangle = 14.38\pm0.18$.  If considered separately, the 
column density distributions differ slightly in the northern and 
southern Galactic skies for high Galactic latitudes ($|b| > 45\deg$), 
with a northern column density enhancement 
of $\sim0.25$ dex above that measured in the south.

The high velocity \ion{O}{6} features have logarithmic column densities
of 13.06 to 14.59, with an average of 
$\langle \log N \rangle = 13.95\pm0.34$.   Typically, the 
high velocity \ion{O}{6} accounts for 25-40\% of the total 
\ion{O}{6} column density along sight lines for which both
have been measured.  The column density
distributions for the Galactic halo and high velocity clouds are 
shown in the left panel of Figure~3.

\begin{figure}[ht!]
\includegraphics{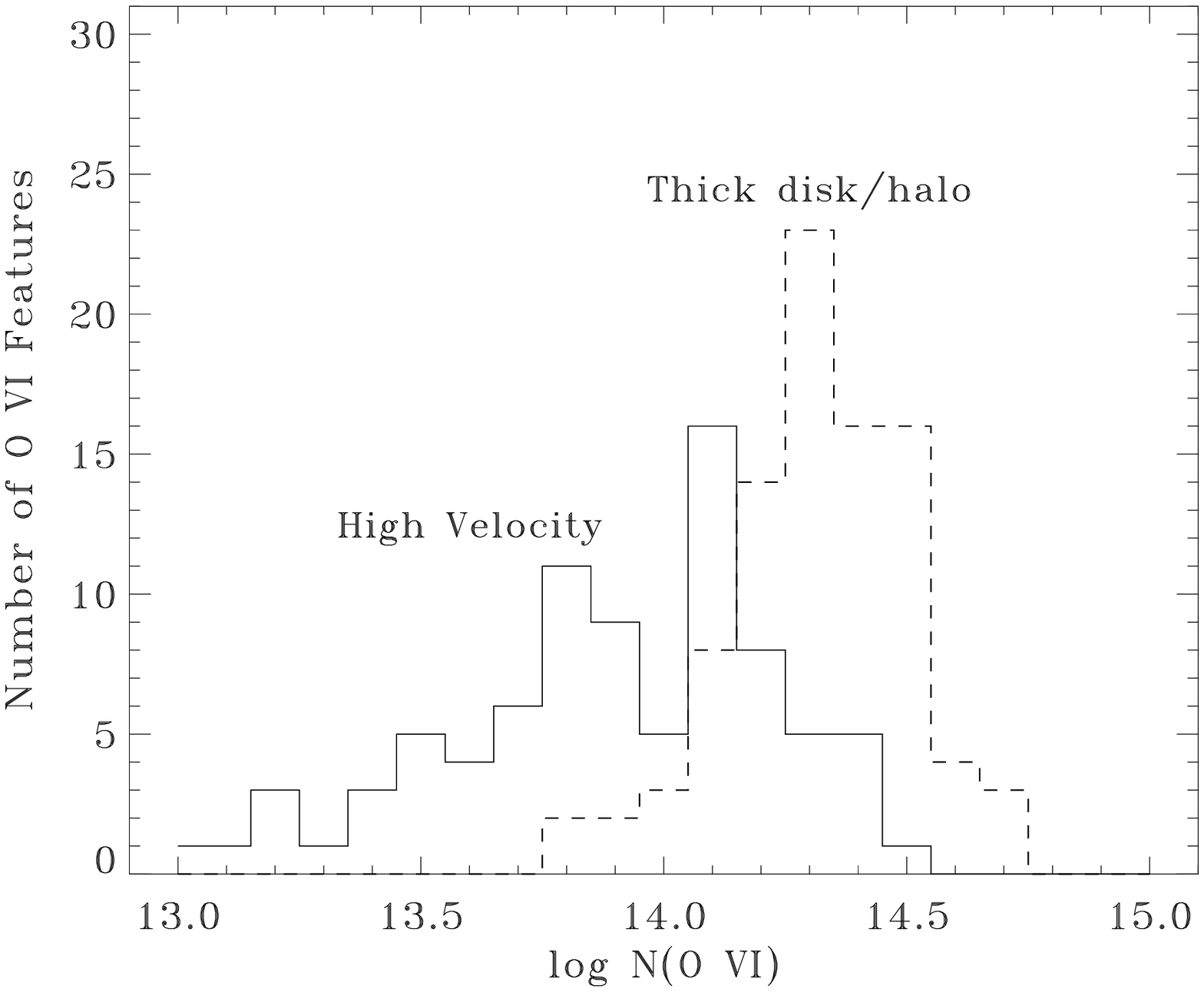}
\includegraphics{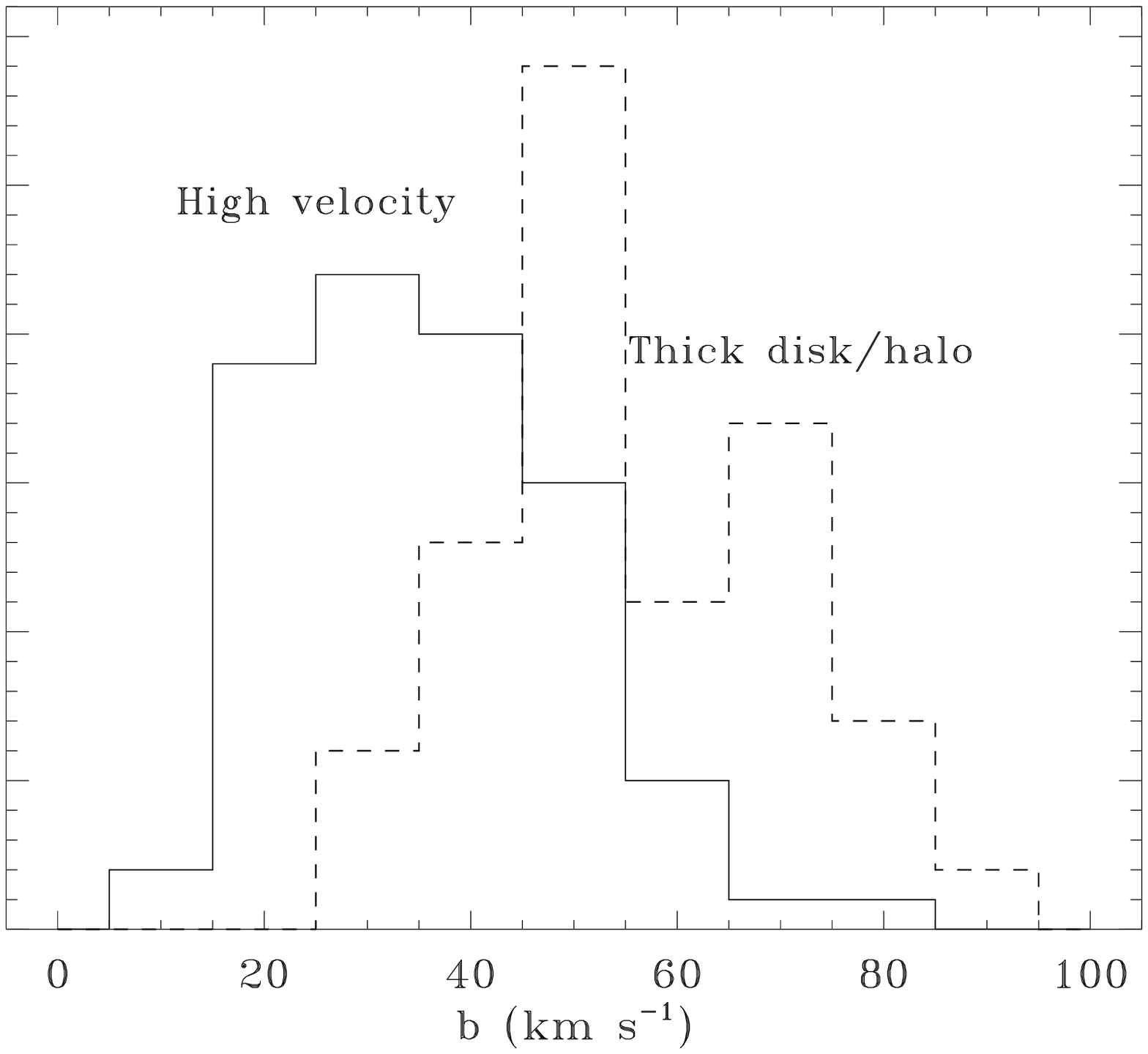}
\vspace{2.0in}
\caption{\footnotesize
Histograms of the logarithmic O\,{\sc vi} column densities and Doppler
line
widths for the Galactic thick disk/halo (dashed lines) and high
velocity clouds (solid lines). The bin sizes are 0.10 dex and 10 \kms, 
respectively.}
\end{figure}

\subsection{Line Widths}

The line widths of the  Galactic halo \ion{O}{6} features range from
$\sim$30 km~s$^{-1}$ to $\sim$99 km~s$^{-1}$, with an average of 
$\langle {\rm b} \rangle = 60\pm15$ km~s$^{-1}$.
The line widths of the high velocity \ion{O}{6} features range from
$\sim$16 km~s$^{-1}$ to $\sim$81 km~s$^{-1}$, with an average of 
$\langle {\rm b} \rangle = 40\pm14$ km~s$^{-1}$.
The lowest values of b in the high velocity gas 
are close to the thermal width of 17.1 km~s$^{-1}$ expected for \ion{O}{6}
at its peak ionization fraction temperature of $T = 2.8\times10^5$\,K in
collisional ionization equilibrium (Sutherland \& Dopita 1993).
The large b-values observed along many sight lines imply that non-thermal broadening mechanisms and multi-component velocity structure dominate the \ion{O}{6} line broadening in both the halo gas and
in the high velocity clouds.  The distributions of b-values
for both types of gas are shown in the right panel of Figure~3.

\section{\ion{O}{6} in the Galactic Thick Disk/Halo}

\subsection{Distribution}
Strong \ion{O}{6} absorption arising in the low Galactic halo 
is measurable in 91 of the 102 sight lines in our survey. In
those cases where it was not detected, the upper limits on N(\ion{O}{6})
are generally higher than the smallest column density measured in the 
sample.  Thus, the apparent absence of \ion{O}{6} along some sight lines
is probably an artifact of the varying data quality within the sample.
FUSE observations of  halo stars within a few
kiloparsecs of the Galactic plane  reveal
Galactic halo \ion{O}{6} absorption 
at a column density level consistent with approximately 40--50\% of the 
halo absorption seen along complete halo paths (Zsarg\'o et al.\ 2003).

The distribution of Galactic halo \ion{O}{6} is patchy, which suggests 
that the \ion{O}{6} may be distributed in filamentary or small-scale
structures.  The \ion{O}{6} regions must have a large covering factor 
to account for the common occurrence of \ion{O}{6} absorption in
halo directions, but need not have a large volume filling factor.
Variations of a factor 
of three in N(\ion{O}{6}) have been seen across the face of the LMC at 
angular scales of $\sim1\deg$ (Howk et al.\ 2002a). These variations are 
comparable to those seen over much larger angular separations in the 
FUSE survey.   Smaller variations 
have been seen toward both the LMC and SMC at smaller angular 
separations (Howk et al.\ 2002b; Hoopes et al. 2002).
Measurements of \ion{O}{6} in the direction of the globular cluster 
NGC\,6752 (Lehner \& Howk 2004) show no variations in N(\ion{O}{6})
or the \ion{O}{6} 
velocity distribution on angular scales $\Delta\theta = 2\farcm2-8\farcm9$
(or 2.5--10.1 pc).  The constancy of the \ion{O}{6} absorption in
this direction suggests that there may be a characteristic size of $\ga 10$ 
parsecs
for the \ion{O}{6} regions, but similar checks for several other directions 
need to be conducted before such a conclusion can be generalized to
the entire Galactic halo.

Nearby regions of hot gas do not appear to contribute significantly to 
the \ion{O}{6} variations.  There is no obvious enhancement in N(\ion{O}{6})
in the directions of Radio Loops I-IV, with the possible exception of 
the 3C\,273 sight line, which passes through a region of enhanced 
soft X-ray emission in the North Polar Spur region (see Sembach et al.\ 2001
and Figure 9 of Savage et al.\ 2003).  Similarly, the Local Bubble 
contains only a small amount of \ion{O}{6} (typically, log~N $\la 13.0$;
Oegerle et al.\ 2004) and 
contributes at most $\sim10\%$ of the \ion{O}{6} column along the extended
paths probed in the AGN/QSO survey.

\begin{figure}[h!]
\includegraphics{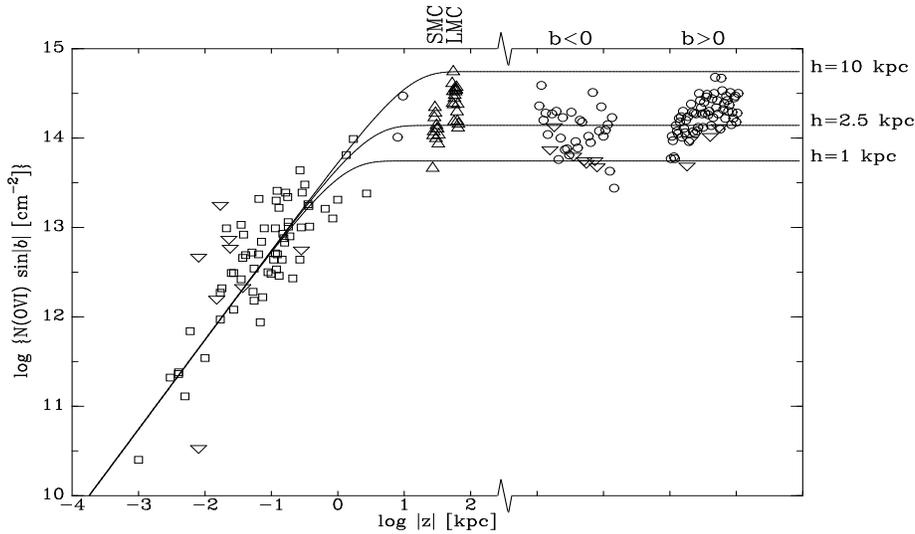}
\vspace{2.8in}
\caption{\footnotesize Extension of O\,{\sc vi} away from the Galactic plane.  
The data points shown include the FUSE  AGN/QSO 
survey (open circles), the {\it Copernicus} disk sample (squares), and the 
FUSE LMC and SMC samples (open upward triangles).  Downward triangles indicate
3$\sigma$ upper limits within each sample. The best fit exponentially 
distributed gas layer
for the data points shown has a scale height of $h_{\rm O\,VI} = 2.3$ 
kpc for a mid-plane density n$_0$(O\,{\sc vi})$ = 1.7\times10^{-8}$ cm$^{-3}$.
From Savage et al.\ (2003).}
\end{figure}

It is possible to estimate the extension of the \ion{O}{6} gas distribution
away from the Galactic plane by examining the column density
perpendicular to the plane (N sin$|b|$) as a function of distance of the 
background
source from the plane (z). Figure~5 shows this distribution for the survey
sight lines as well as sight lines toward the Magellanic Clouds and 
stars in the Galactic disk.  The solid curves on the plot show the 
predicted relation between N$_\perp$ = N sin$|b|$ and z for a smooth,
 exponentially distributed
gas layer with a mid-plane density 
n$_0$(O\,{\sc vi})$ = 1.7\times10^{-8}$ cm$^{-3}$ and scale heights
of 1, 2.5, and 10 kpc.  The north-south column density difference mentioned 
in \S2.2 is evident in this plot. The simplest model for the gas is 
one with an exponential scale height of 2.3 kiloparsecs with log~N$_\perp = 
14.09$ and a 0.25 dex 
enhancement in N$_\perp$ at high positive Galactic latitudes.
This 
distribution is roughly consistent with those for other high ionization 
species such as \ion{Si}{4}, \ion{C}{4}, and \ion{N}{5} ($h\sim 3-5$ kpc;
Savage, Sembach, \& Lu 1997).  A careful comparison of the \ion{O}{6}
and other high ions along many individual sight 
lines is needed to fully assess the relative kinematics and 
distributions of these species (see, e.g., Zsarg\'o et al.\
2003 and Indebetouw \& Shull 2004).

\subsection{Relationship to Other Gas Tracers}

The distribution and total amount of \ion{O}{6} generally does not correlate 
well with other tracers of ionized interstellar gas, such as soft x-ray 
emission 
from gas with $T \sim 10^6$\,K or H$\alpha$ emission from the warm 
($T\sim10^4$\,K) ionized ISM.  Similarly, N(\ion{O}{6})
 does not track the total intensity of \ion{H}{1} 21\,cm emission.  However,
it is common for \ion{O}{6} to be found at velocities similar to those
of \ion{H}{1} and lower ionization species, such as \ion{C}{2-III-IV} and
\ion{Si}{2-III-IV}.  This result
suggests that the \ion{O}{6} arises in distinct regions that are 
kinematically coupled to the lower ionization gas, such as interfaces
between warm and hot gas.

\subsection{Production of the Hot Gas}

In the Milky Way, it is difficult to convert \ion{O}{5} into \ion{O}{6}
through photoionization.  The ambient interstellar radiation field and
the starlight from all but the hottest He-poor stars lack the 114\,eV 
photons necessary for this conversion.  Furthermore, photoionization by
the extragalactic ultraviolet background is also inefficient at producing
\ion{O}{6} in the local universe; path lengths of hundreds of 
kiloparsecs are required to reproduce the observed \ion{O}{6} columns
(see Figure~20 in Sembach et al.\ 2003).   As a result, the vast 
majority of the \ion{O}{6} observed along extended paths through 
the Galactic halo must be created by collisional processes.
\ion{O}{6} is an excellent tracer of collisionally ionized gas at temperatures
of $(1-5)\times10^5$\,K. 

A variety of processes have been suggested as possible high ion production
mechanisms (turbulent mixing, conduction, radiative cooling, shocks, etc).
It is not possible to describe all of these here.  For reviews, see the 
discussions in Fox et al.\ (2004) and  Sembach, Tripp, \& Savage (1997).
A combination of models involving radiatively cooling gas, turbulent mixing, 
and the cooling of superbubbles in the Galactic halo may be necessary to
explain all of the \ion{O}{6} data.  If the origin of the \ion{O}{6} is
dominated by cooling gas in a Galactic fountain flow, Savage et al.\ (2003) 
estimate that
a mass flow rate of 1.4 $M_\odot$ yr$^{-1}$ to each side of the Galactic 
plane with an average density of $10^{-3}$ cm$^{-3}$ is required 
to explain the  
the average value of N$_\perp$(\ion{O}{6}) = $1.2\times10^{14}$ cm$^{-2}$  
measured for the 
southern Galactic hemisphere.

\section{High Velocity \ion{O}{6}}

\subsection{Sky Covering Fraction}
Sembach et al.\ (2003) have identified 84 individual 
high velocity \ion{O}{6} features along 102 sight lines observed by FUSE. 
With few exceptions, the high velocity \ion{O}{6}
absorption is confined to $100 \le |v_{LSR}| \le 400$ km~s$^{-1}$,  
indicating that the identified \ion{O}{6} features are either associated with 
the Milky Way or are nearby clouds within the Local Group.  

We detect high velocity \ion{O}{6} $\lambda1031.926$ absorption 
with total equivalent widths $W_\lambda > 30$ m\AA\ ($3\sigma$)
 along 59 of the 102 sight lines surveyed.  For the highest 
quality sub-sample of the dataset, the high velocity detection frequency 
increases to 22 of 26 sight lines.  Forty of the 59 
sight lines have  high velocity \ion{O}{6} $\lambda1031.926$ absorption
with  $W_\lambda > 100$ m\AA, and 27 have total equivalent widths 
$W_\lambda > 150$~m\AA.  
Converting these \ion{O}{6} equivalent width detection frequencies
 into estimates of $N$(H$^+$) in the 
hot gas indicates
that $\sim60$\% of the sky (and perhaps as much as $\sim85$\%) 
is covered by hot ionized hydrogen at a level of 
\begin{equation}
N({\rm H}^+)\ga 10^{18}~ \left(\frac{f_{\rm O\,VI}}{<0.2}\right)\left(\frac{Z}{0.2Z_\odot}\right) {\rm cm}^{-2},
\end{equation}
where $f_{\rm O\,VI}$ is the fraction of oxygen in the form of \ion{O}{6},
and $Z/Z_\odot$ is the gas metallicity relative to the Sun.  
This H$^+$ detection frequency is
 larger than the value of $\sim37$\% found for warm, high velocity neutral 
gas with 
$N$(\ion{H}{1})\,$\sim 10^{18}$ cm$^{-2}$ traced through 21\,cm emission
(Lockman et al.\ 2002).

\subsection{Distribution}

Figure~4 shows the locations of the high velocity \ion{O}{6} features on the 
sky, with squares indicating positive velocity features and circles 
indicating negative velocity features.  The magnitude of the LSR velocity 
(without sign) is given inside each square or circle for the dominant
high velocity \ion{O}{6} feature along each sight line.  For sight lines
with multiple high velocity features, we have plotted the information 
for the feature with the greatest \ion{O}{6} column density.   
Solid dots indicate directions where no high velocity \ion{O}{6} was detected.

\begin{figure}[h!]
\includegraphics{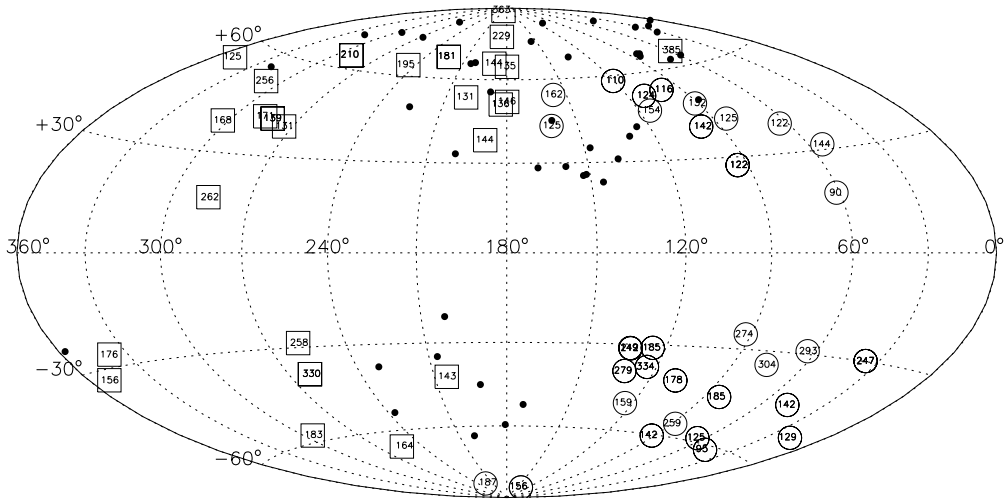}
\vspace{2.41in}
\caption{\footnotesize High velocity O\,{\sc vi} 
features along the sight lines in the FUSE O\,{\sc vi} survey.  Squares
indicate positive velocity features.  Circles indicate negative 
velocity features.  The magnitude of the velocity is given inside each 
circle or square for the highest column density O\,{\sc vi} HVC component 
along each sight line.  Solid dots indicate sight lines with no high 
velocity O\,{\sc vi} detections.}
\end{figure}

There is often good correspondence between high velocity 
\ion{H}{1} 21\,cm emission and high velocity \ion{O}{6} absorption.
When high velocity \ion{H}{1} 21\,cm emission
is detectable in a particular direction,
high velocity \ion{O}{6} absorption is usually detected if a suitable 
extragalactic continuum source is bright enough for FUSE to observe.   
\ion{O}{6} is present in the Magellanic Stream, which passes
through the south Galactic pole and extends up to $b \sim -30^o$,
with positive velocities for $l \ga 180^o$ and
negative velocities for $l \la 180^o$.
\ion{O}{6} is present in high velocity cloud Complex~C, which
covers a large portion of the
northern Galactic sky between $l=30^o$ and $l = 150^o$ and has
velocities of roughly --100 to --170 \kms.

The segregation of positive and negative velocities in Figure~4 is striking, 
indicating that the clouds and the underlying (rotating) disk of the Galaxy 
have very different kinematics.  A similar velocity pattern is seen for 
high velocity \ion{H}{1} 21\,cm emission and has sometimes been used to 
argue for a Local Group location for the high velocity clouds (see Blitz
et al.\ 1999).  The kinematics of the high velocity
\ion{O}{6} clouds are consistent with a distant location, but do not 
necessarily require an extended Local Group distribution as proposed by
Nicastro et al.\ (2003). The dispersion about the 
mean of the high velocity \ion{O}{6}
centroids  decreases when the velocities are converted from the
Local Standard of Rest (LSR) into the Galactic Standard of Rest (GSR) and 
the Local Group Standard of Rest (LGSR) reference frames with various 
assumptions about the distances of the clouds and their transverse 
motions.   While this 
reduction is expected if the 
\ion{O}{6} is associated with gas in a highly extended Galactic corona,  
it {\it does not} provide sufficient proof by itself of an
extended extragalactic distribution for the high velocity gas because the
correction to the LGSR reference frame requires proper knowledge of the total
space velocities of the clouds.  Only one component of motion - the 
velocity toward the Sun - is observed.  Additional information,
such as the gas metallicity, ionization state, or parallax resulting from
transverse motion 
across the line of sight, is needed to constrain 
whether the clouds are located near the Galaxy or are farther away in the 
Local Group.

\subsection{Positive Velocity Absorption Wings}

Twenty-one of the sight lines in the FUSE survey exhibit broad,
weak absorption wings at high positive velocities.  These wings extend
asymmetrically from low velocities, representative of those of Galactic
halo gas, out to much higher velocities.  In most cases, the integrated
\ion{O}{6} column densities in the broad
wings are much less than N(\ion{O}{6}) in the lower velocity absorption.
Of the 21 sight lines with broad wings, 
17 are in the northern Galactic hemisphere.

Some of the wings may be associated with outflows of hot gas from the 
Galactic disk, as has been suggested for the positive velocity 
wing in the direction of 3C\,273 (Sembach et al.\ 2001). Toward
Mrk\,421, there is no corresponding wing of absorption in two nearby 
halo stars, implying that the wing arises in gas at distances of more than
3.5 kiloparsecs from the Galactic plane (Savage et al.\ 2005).  This result
may hold more generally as well, since no such wings are observed toward 
a larger sample of halo stars spread across the sky (Zsarg\'o et al.\ 2003).
Some of the wing gas could be tidal debris from past encounters 
between the Milky Way and Magellanic Clouds or other dwarf galaxies.
A thorough study of the \ion{O}{6} wing absorption features and their
relationship (or lack thereof) to lower ionization absorption may help
to reveal their origin.

\subsection{Origin of the High Velocity \ion{O}{6}}

One possible explanation for some of the high velocity \ion{O}{6} is that
transition temperature gas arises at the boundaries between cool/warm 
clouds of gas and a very hot ($T > 10^6$\,K) Galactic corona or Local Group
medium.   Sources of the high velocity material might include infalling or 
tidally disturbed galaxies.  
 A hot, highly extended ($R > 70$ kpc)
corona or Local Group medium might be left over from the formation of the 
Milky Way or Local Group, or may be the result of continuous accretion of 
smaller galaxies over time.  
Hydrodynamical simulations of clouds moving through a hot, low-density 
medium show that weak bow shocks develop on the leading edges of the 
clouds as the gas is compressed and heated (Quilis \& Moore 2001).  
Even if the clouds are 
not moving at 
supersonic speeds relative to the ambient medium, turbulent
stripping of the cooler gas should result in mixing with the hotter
coronal gas.  \ion{O}{6} would be produced in the transition temperature
gas resulting from the mixing.  This mechanism for producing \ion{O}{6}
has been suggested for Complex~C (Fox et al.\ 2004) as well as smaller
isolated high velocity clouds (e.g., Ganguly et al.\ 2005).
Additional evidence for a hot Galactic corona through which the high
velocity clouds must pass comes from the detection 
of \ion{O}{7} X-ray absorption lines near zero velocity
(e.g., Nicastro et al.\ 2002;
Fang, Sembach, \& Canizares 2003; 
Rasmussen et al.\ 2003; Futamoto et al.\ 2004;
McKernan, Yaqoob, \& Reynolds 2004).

Alternatively, the \ion{O}{6} high velocity clouds 
 and any associated \ion{H}{1} or lower ionization material  may be  
condensations within
large gas structures falling onto the Galaxy.
Cosmological structure formation models predict large numbers of cooling 
fragments embedded in dark matter, and some of these structures should be 
observable in \ion{O}{6} absorption as the gas cools through the 
$T=10^5-10^6$\,K
temperature regime (Dav\'e et al. 2001). 

Some of the high velocity \ion{O}{6} clouds may be extragalactic 
clouds, based on what we currently know about their ionization 
properties (see, for example, Sembach et al.\ 1999).  However, claims that
essentially  {\it all} of the \ion{O}{6} HVCs are extragalactic 
entities associated with an extended Local Group filament based on
kinematical arguments alone are untenable.
 Such arguments
fail to consider the selection biases inherent in the \ion{O}{6} 
sample, the presence of neutral (\ion{H}{1}) and lower ionization 
(\ion{Si}{4}, \ion{C}{4})
gas associated with some of the \ion{O}{6} HVCs, and the known ``nearby''
locations for at least two of the primary high velocity complexes 
in the sample ---  the Magellanic Stream is circumgalactic tidal debris, and 
Complex~C is interacting with the Galactic corona (see Fox et al.\
2004; Collins, Shull, \& Giroux 2003).  Furthermore, the  
O\,{\sc vii} X-ray absorption measures used to support an extragalactic 
location have not yet been convincingly tied to either the \ion{O}{6} HVCs
or to an external Local Group location.  The O\,{\sc vii} absorption may 
well have a 
significant Galactic component in some directions
(see Fang et al.\ 2003). The Local Group filament 
interpretation (Nicastro et al.\ 2003) may be suitable for some of the 
observed high velocity \ion{O}{6} features, but it clearly fails in 
other cases (e.g., the Magellanic Stream or the PKS\,2155-304 HVCs -- see
Sembach et al.\ 2003 and Collins, Shull, \& Giroux 2004).


\begin{references}
\reference{}Blitz, L., et al.\  1999, ApJ, 514, 818
\reference{}Collins, J.A., Shull, J.M., \& Giroux, M.L.\  2003, ApJ, 585, 336
\reference{}Collins, J.A., Shull, J.M., \& Giroux, M.L.\  2004, ApJ, 605, 216
\reference{}Dav\'e, R., et al.\ 2001, ApJ, 552, 473
\reference{}Fang, T., Sembach, K.R., \& Canizares, C.R.\  2003, ApJ, 586, L49
\reference{}Fox, A., et al.\  2004, ApJ, 602, 738
\reference{}Futamoto, K., et al.\ 2004, ApJ, 605, 793
\reference{}Ganguly, R., Sembach, K.R., Tripp, T.M., \& Savage, B.D.  2005,
	ApJ, in press [astro-ph/0412485]
\reference{}Hoopes, C.G., et al.\ 2002, ApJ, 569, 233
\reference{}Howk, J.C., Savage, B.D., Sembach, K.R., \& Hoopes, C.G.\
	2002a, ApJ, 572, 264
\reference{}Howk, J.C., et al.\ 2002b, ApJ, 569, 214
\reference{}Indebetouw, R., \& Shull, J.M.\  2004, ApJ, 607, 309
\reference{}Lehner, N., \& Howk, J.C.\  2004, PASP, 116, 895
\reference{}Lockman, F.J., Murphy, E.M., Petty-Powell, S., \& Urick, V.\
        2002, ApJS, 140, 331
\reference{}McKernan, B., Yaqoob, T., \& Reynolds, C.S.\ 2004, ApJ, 617, 232
\reference{}Nicastro, F., et al. 2002, ApJ, 573, 157
\reference{}Nicastro, F., et al.\ 2003, Nature, 421, 719
\reference{}Oegerle, W.R., et al.\ 2004, ApJ, submitted [astro-ph/0411065]
\reference{}Quilis, V., \& Moore, B.  2001, ApJ, 555, L95
\reference{}Rasmussen, A., Kahn, S.M., \& Paerels, F.  2003, in the 
        IGM/Galaxy Connection, eds. J.L. Rosenberg
        \& M.E. Putman, (Dordrecht: Kluwer), 109 [astro-ph/0301183]
\reference{}Savage, B.D., Sembach, K.R., \& Lu, L.\  1997, AJ, 113, 2158
\reference{}Savage, B.D., et al.\ 2003, ApJS, 146, 125
\reference{}Savage, B.D., et al.\ 2005, ApJ,
	in press [astro-ph/0412381]
\reference{}Sembach, K.R., Savage, B.D., Lu, L., \& Murphy, E.  1999, ApJ,
	515, 108
\reference{}Sembach, K.R., et al.\  2001, ApJ, 561, 573
\reference{}Sembach, K.R., Tripp, T.M., \& Savage, B.D.  1997, ApJ, 480, 216
\reference{}Sembach, K.R., et al.\ 2003, ApJS, 146, 165
\reference{}Sutherland, R.S., \& Dopita, M.A.\  1993, ApJS, 88, 253
\reference{}Wakker, B.P., et al. 2003, ApJS, 146, 1
\reference{}Zsarg\'o, J., Sembach, K.R., Howk, J.C., \& Savage, B.D.\
	2003, ApJ, 586, 1019
\end{references}
\end{document}